\documentclass[preprint]{imsart}

\RequirePackage{amsthm,amsmath,amsfonts,amssymb}
\RequirePackage[authoryear]{natbib}

\RequirePackage[colorlinks,citecolor=blue,urlcolor=blue,backref=page,backref=page]{hyperref}
\RequirePackage{graphicx}
\RequirePackage{float}
\RequirePackage{appendix}

\pubyear{2024}
\volume{TBA}
\issue{TBA}
\firstpage{1}
\lastpage{1}

\startlocaldefs
\theoremstyle{plain}

\theoremstyle{definition}

\theoremstyle{remark}

\def\giturl{https://github.com/mathematica-mpr/bcf-1}
\endlocaldefs

\begin{document}

\begin{frontmatter}
\title{Aggregate Bayesian Causal Forests: The ABCs of Flexible Causal Inference for Hierarchically Structured Data}
\runtitle{Aggregate Bayesian Causal Forests}

\begin{aug}
\author[A]{\fnms{Dan R. C.}~\snm{Thal}\ead[label=e1]{dthal@mathematica-mpr.com}},
\author[A]{\fnms{Lauren V.}~\snm{Forrow}\ead[label=e2]{lforrow@mathematica-mpr.com}},
\author[B]{\fnms{Erin R.}~\snm{Lipman}\ead[label=e3]{erlipman81@gmail.com}},
\author[A]{\fnms{Jennifer E.}~\snm{Starling}\ead[label=e4]{jstarling@mathematica-mpr.com}},
\and
\author[C]{\fnms{Mariel M.}~\snm{Finucane}\ead[label=e5]{Mariel.Finucane@bcbsma.com}}
\address[A]{Mathematica\printead[presep={,\ }]{e1,e2,e4}}
\address[B]{Department of Statistics, University of Washington\printead[presep={,\ }]{e3}}
\address[C]{Blue Cross Blue Shield of Massachusetts\printead[presep={,\ }]{e5}}
\runauthor{Thal et al.}
\end{aug}

\begin{abstract}
This paper introduces aggregate Bayesian Causal Forests (aBCF), a new Bayesian model for causal inference using aggregated data. Aggregated data are common in policy evaluations where we observe individuals such as students, but participation in an intervention is determined at a higher level of aggregation, such as schools implementing a curriculum. Interventions often have millions of individuals but far fewer higher-level units, making aggregation computationally attractive. To analyze aggregated data, a model must account for heteroskedasticity and intraclass correlation (ICC). Like Bayesian Causal Forests (BCF), aBCF estimates heterogeneous treatment effects with minimal parametric assumptions, but accounts for these aggregated data features, improving estimation of average and aggregate unit-specific effects. 

After introducing the aBCF model, we demonstrate via simulation that aBCF improves performance for aggregated data over BCF. We anchor our simulation on an evaluation of a large-scale Medicare primary care model. We demonstrate that aBCF produces treatment effect estimates with a lower root mean squared error and narrower uncertainty intervals while achieving the same level of coverage. We show that aBCF is not sensitive to the prior distribution used and that estimation improvements relative to BCF decline as the ICC approaches one. Code is available at 
 \url{\giturl}.
\end{abstract}

\end{frontmatter}

\begin{acks}[Acknowledgments]
	The authors gratefully acknowledge Peter Mariani and Constance Delannoy for their help implementing the heteroskedastic errors model and organizing the codebase. We thank Laura Blue, Nancy McCall, John Deke, and Dave Jones for providing ideas that shaped the paper and thoughtful feedback on the draft. Finally, we thank Tim Day and Georgia Pearson for their unflagging support of this work and their helpful comments on early drafts of this paper. This work was funded by the U.S. Department of Health and Human Services, Center for Medicare and Medicaid Services Innovation Center under contract HHSM-500-2014-0034I/75FCMC19F0005.

	Disclaimer: The contents of this manuscript are solely the responsibility of the authors and do not necessarily represent the official views of the U.S. Department of Health and Human Services or any of its agencies. 
\end{acks}

\section{Introduction}

Analysis of aggregate observational data is common in many settings, including health policy evaluations, studies of education reform, and analysis of geographic phenomena such as local temperatures. In these settings, assigning interventions at the individual level is often infeasible (for example, individual students in the same classroom cannot use different curricula), so assignment to an intervention occurs at the aggregate level (for example, classroom or school). Further complicating the analysis is that assignment to an intervention is typically voluntary (for example, a school decides whether to adopt a new curriculum). Even in these settings, we still assess effectiveness by measuring individuals’ outcomes, such as individual students’ test scores. To accurately estimate an intervention’s effect in these settings, it is important to be able to model individual outcomes, reflect the aggregated nature of the intervention, and account for confounding bias that could arise from voluntary assignment of aggregate units to an intervention. Failure to account for these features risks misinterpretation of selection effects as causal effects and can yield overly optimistic uncertainty intervals.

Analyzing aggregated data has practical and conceptual advantages. From a practical point of view, aggregated data sets dramatically reduce sample sizes, lowering computational burden. For large-scale interventions such as nationwide Medicare policies (Medicare is a U.S. federal health insurance program for those age 65 or older, or with certain disabling conditions), there might be millions of individuals (for example, Medicare beneficiaries) but only thousands of aggregate units (for example, primary care practices). For more advanced and computationally intensive approaches, such as techniques that rely on Bayesian methods to stabilize estimates for small subgroups, that flexibly model effect heterogeneity, or that provide intuitive probabilistic quantification of uncertainty, it is impractical and often infeasible to analyze individual-level data directly. To leverage these advanced techniques, researchers must analyze aggregated data.

Conceptually, it is important to model intervention effects at the level of assignment to the intervention \citep{deke2016, abadie2017}. When aggregate units such as primary care practices or schools volunteer for an intervention, confounding because of voluntary participation arises at the aggregate level. Because all individuals in the same aggregate unit necessarily have the same treatment status, there is no within-unit variation in treatment, so within-unit variation in covariates cannot confound. Researchers can account for confounding by including aggregate measures of individual covariates, such as the proportion of a primary care practice’s patients that are female, and aggregate unit-level measures, such as whether a practice operates independently, in their models \citep{thal2023}.

In addition to the main task of estimating an intervention’s effects, policy evaluations are often interested in understanding the overall performance of each unit, regardless of whether that performance is because of treatment or would have occurred regardless. For example, “value-added” models are common in education; these models use teacher-level random effects to measure teacher and school performance \citep{kane2008, chetty2014}. In evaluations of health care policies, identification of exemplars (that is, practices or providers doing well above and beyond what can be expected based on their covariates and patient mix) are of similar interest \citep{swankoski2022}. 

We propose a new method called aggregate Bayesian Causal Forests (aBCF) to estimate causal effects using aggregated data. Our method builds on Bayesian Causal Forests (BCF) \citep{hahn2020}, whose flexible nonparametric approach is well equipped to model heterogeneous treatment effects and account for confounding. aBCF additionally accounts for aggregate data’s heteroskedasticity and intraclass correlations and allows users to estimate aggregate units’ individual outcomes.

The paper proceeds as follows. In Section~\ref{secbayestree}, we review existing Bayesian tree models used for causal inference. In Section~\ref{secmethod}, we introduce aBCF and describe its innovations. In Section~\ref{secsim}, we introduce the simulation framework we use to test performance and compare aBCF to BCF. Lastly, in Section~\ref{secdiscuss}, we discuss limitations and next steps.

\section{Bayesian tree models for causal inference} \label{secbayestree}
In this section, we review the BCF model and Bayesian Additive Regression Trees (BART) \citep{chipman2010}, the model upon which BCF is built. BART is a non-parametric regression model that estimates conditional outcomes via a sum of binary decision trees. The sum-of-trees approach allows BART to flexibly estimate functions without any parametric assumptions, and the Bayesian priors regularize to guard against overfitting. For response $y_i$, a scalar outcome, and $x_i$, a vector of covariates, BART models $y_i$ as 
\[y_i = f(x_i)+\varepsilon_i\]
\[\varepsilon_i \sim N(0,\sigma_\varepsilon^2)\]
where $f(x_i)$ is the sumn of tree-specific mapping functions $g(.)$ such that:
\[f(x_i) = \sum_{k=1}^{m}g(x_i; T_k,M_k)\]
In this model, $T_k$ is a binary tree that separates the covariate space via binary splits in $x_i$, and $M_k={\mu_{1k}…\mu_{bk}}$ are the $b_k$ different leaf node values for the $k$th tree. We refer readers to \citet{chipman1998} and \citet{chipman2010} for details about the BART priors.
BART has demonstrated powerful predictive performance \citep{chipman2010, hill2020} and has been used for causal inference for more than a decade. By fitting a BART model to the outcome of interest, including treatment status ($z_i$) as a covariate, and then taking the difference in predicted outcomes with treatment status set to 1 and 0, we can estimate treatment effects, $\tau_i$, for any combination of covariates represented by $x_i$ \citep{hill2011}:
\[y_i=f(x_i,z_i )+\varepsilon_i\]
\[\tau_i=f(x_i,z_i=1)-f(x_i,z_i=0)\]
Used in this way, BART has performed well in multiple instances of the American Causal Inference Conference blinded causal inference competitions \citep{dorie2019, thal2023}. These competitions pit novel causal inference methodologies against each other on a series of simulated data sets, providing an even playing field for judging performance across techniques because none of the competitors know the data generating procedure.
BCF tailors BART for causal inference by making two additional changes.
\[y_i=\mu(x_i,\hat{\pi}_i )+z_i \tau(x_i )+\varepsilon_i\]
\[\varepsilon_i  \sim N(0,\sigma_\varepsilon^2)\]
First, BCF separately models two components of the outcome using BART functions: the first, $\mu(x)$, models counterfactual outcomes, and the second, $\tau(x)$, models treatment effects. With this setup, we can directly interpret $\tau$ as the causal estimate. Separately modeling counterfactual outcomes and treatment effects also allows for prior distributions tailored for each component. \citet{hahn2020} place a more stringent prior on the $\tau$ trees than the $\mu$ trees, allowing the model to shrink back toward homogeneous treatment effects without undermining its ability to represent complex confounding relationships.

Second, BCF incorporates $\hat{\pi}_i$, an estimate of the propensity score, the probability that a unit joins the treatment group (mathematically, $\pi_i=Pr(z_i=1|x_i)$), as a covariate in $\mu$. Including a univariate summary of confounding relationships is especially beneficial because the BART model’s preference for simplicity might inhibit adjustments for complex confounding, a phenomenon called regularization-induced confounding \citep{hahn2020}. The propensity score offers the model a shortcut to adjust for confounding without violating its preference for parsimony. BCF has shown improvements over BART for causal inference in American Causal Inference Conference challenges, with somewhat smaller uncertainty interval widths and better estimation of unit-specific effects \citep{hahn2020}.

Although BCF offers distinct advantages over BART for causal inference, it lacks features that allow it to analyze aggregated data, which is notably limiting in its application to program and policy evaluations that commonly feature aggregate units, such as primary care practices or schools. \citet{yeager2019} extended BCF for their analysis to incorporate weights, but these were survey sampling weights, and the study still used individual data, so these weights fulfilled a different purpose from ours.

\section{Methods} \label{secmethod}
In this section, we start by expanding our notation and describing our assumptions. We then define the mathematical model underpinning aBCF. We provide the code for running aBCF in the abcf R package available on online at \url{\giturl}.

\subsection{Notation and assumptions} \label{subsecnotation}
We extend the notation introduced in Section~\ref{secbayestree} to cover the aggregate case. Let $y_{ij}$ denote the response variable for individual $i$ belonging to aggregate unit $j$. Let $z_j$ denote a binary treatment indicator, identical for all $i$ within each $j$, and let $x_j$  be a vector of covariates. In our example, we measure covariates at the aggregate level, although they can include aggregate measures of individual-level covariates, $q_i$. Most commonly, these aggregated covariates are the mean of individual-level covariates, such as the proportion of patients in a given age group at a primary care practice, or some other function such that $x_j=f(q_i)$. Let $w_j$ be the number of individuals belonging to aggregate unit $j$. We presume some variation across units in the number of individuals, as this is necessary for identification of the variance components. In our examples, we discuss some observed sample of $n$ aggregate units, in which for each unit we observe $y_j,z_j,x_j,w_j$. We could instead think of this as having a sample of $\sum w_j$ individuals and, for each observing the aggregate unit to which they belong, $J_i$, their covariates, the covariates of the unit to which they belong, and their outcomes. 

In our causal inference setting, we are interested in identifying $\tau(x_j)$, the treatment effect for aggregate unit $j$, modeled as a function of observed covariates. Using the potential outcomes framework of \citet{imbens2015}, we can think of the treatment effect for any unit of aggregation as the difference between the outcome that occurs under treatment, $y_j(1)$, and the same unit’s outcome without treatment, $y_j(0)$. 

Throughout this paper we make three common assumptions for causal identification. The first is strong ignorability, also sometimes called no unobserved confounding, which states that both potential outcomes are independent of treatment status, conditional on observed covariates:
\[y_j(0),y_j(1)\perp z_j|x_j\]
Next, we assume conditional overlap (that is, that every aggregate unit had a nonzero probability of belonging to the treatment group and a nonzero probability of belonging to the comparison group, conditional on its covariates):
\[0<Pr(z_j=1|x_j)<1\]
Lastly, we make the stable unit treatment values assumption (SUTVA), and assume that there is a single treatment and that potential outcomes for each unit do not depend on the treatment assignments of other units.

\subsection{Model} \label{subsecmodel}
We introduce the aBCF model by building on the BCF model described by \citet{hahn2020}, expanding it to use aggregate data. As in Section~\ref{secbayestree}, the prognostic function $\mu$ estimates the relationship between covariates and counterfactual outcomes, and the treatment effect function $\tau$ captures both the average treatment effect and effect modification because of covariates.
\[y_j=\mu(x_j,\hat{\pi}_j )+z_j \tau(x_j)+\varepsilon_j\]
\[\varepsilon_j  \sim N(0,\sigma_\varepsilon^2)\]
aBCF extends the model in two ways; both extensions improve the model’s ability to accommodate aggregate data. The first change is to the estimation of the error term, $\varepsilon_j$. aBCF allows for heteroskedasticity by treating the variance of the error term as $\frac{\sigma_\varepsilon^2}{w_j}$, rather than assuming a single common variance across all practices of $\sigma_\varepsilon^2$. With this specification, $\sigma_\varepsilon^2$ is the error variance for a single individual within a unit.

The second change is the addition of unit-level residuals $u_j$ with variance $\sigma_u^2$, which is inferred from the data. The model equation for aBCF is:
\[y_j=\mu(x_j,\hat{\pi}_j )+z_j\tau(x_j )+u_j+\varepsilon_j\]
\[ \varepsilon_j \sim N(0,\frac{\sigma_\varepsilon^2}{w_j})\]
\[u_j \sim N(0,\sigma_u^2)\]
This model reflects the clustered nature of the data via the compound residual term, composed of $u_j+\varepsilon_j$. $\varepsilon_j$ can be thought of as the aggregate-level average of all the relevant individual $\varepsilon_i$, each of which is distributed $N(0,\sigma_\varepsilon^2)$. We can think of the total unexplained variance as $\sigma_u^2+\frac{\sigma_\varepsilon^2}{w_j}$ at the aggregate level and $\sigma_u^2+\sigma_\varepsilon^2$ at the individual level; individuals within a unit are independently and identically distributed, but across units they are not because of the shared $u_j$. This formulation naturally leads to an intuitive understanding of how these two variance components are identified from the data, via the equation for the intraclass correlation coefficient (ICC) from multilevel modeling, $\frac{\sigma_u^2}{\sigma_u^2+\sigma_\varepsilon^2}$) \citep{killip2004}. Specifically, we can identify $\sigma_\varepsilon^2$ by estimating how quickly unexplained variance decreases with sample size, and we can identify $\sigma_u^2$ as the limit, as sample sizes increase, of the total unexplained variance. 

This identification strategy also implies a requirement for aBCF: sample sizes must vary across aggregate units. If each unit has the same sample size, errors are homoskedastic because $\frac{\sigma_\varepsilon^2}{w_j}$ is constant across all units, so we cannot disentangle the two sources of error variation. In that case, the identification of $u_j$ versus $\varepsilon_j$ depends entirely on the prior for the relative magnitudes of $\sigma_\varepsilon^2$ and $\sigma_u^2$, as in any Bayesian model in which a parameter is unidentified by the data and model specification.

\subsection{Priors} \label{subsecpriors}
Most of the priors used by aBCF are unchanged from BCF, including the BART priors on the tree structure and leaf node means for the $\mu$ and $\tau$ functions as well as the priors for various internal scale terms, such as $b_0$ and $b_1$ from \citet{hahn2020}. aBCF slightly modifies the calibration process for the $\sigma_\varepsilon^2$ prior to reflect its new weighted interpretation. Where \citet{chipman2010} use the observed residual variation from a linear regression to ground the prior for $\sigma_\varepsilon^2$, we use this variation to ground the prior on $\frac{\sigma_\varepsilon^2}{\bar{w}}$, where $\bar{w}$ is the average sample size across aggregate units (see \citet{chipman2010} for details on the calibration procedure). The sole new prior aBCF requires is for $\sigma_u$, the scale governing the distribution of $u_j$. For $\sigma_u$ we use a default half-normal prior distribution with scale parameter set to two-thirds of the weighted standard deviation of the observed $y_j$. As prior sensitivity in Section~\ref{subsubsecpriorsens} shows, we do not find that the model is sensitive to this prior.

\section{Simulation Study} \label{secsim}
We explore the performance of aBCF using a simulation study calibrated so that the range of aggregate unit sizes and variability of the outcome resemble a large-scale evaluation of a Medicare primary care model, an application that motivated the development of aBCF. We outline the data generating process, treatment effect estimation, and performance metrics, and then discuss the performance of aBCF relative to BCF for estimating treatment effects and identification of exemplars. Finally, we explore sensitivity to the prior on $\sigma_u$ and how relative performance changes with the ICC.

\subsection{Data generating process (DGP)} \label{subsecdgp}
For each iteration of our simulation, we create a data set containing 3,000 aggregate units covering around 2 million individuals. We simulate the number of individuals comprising each unit and generate each unit’s covariates. Finally, we assign units to the treatment or comparison groups and generate each unit’s outcome.

Each aggregate unit is assigned a number of individuals $w_j$, drawn from the empirical quantiles of the number of Medicare beneficiaries served by primary care practices in the evaluation we use for calibration \citep{swankoski2022}. $w_j$ ranges from 60 to 3,500, with a mean of 621, a median of 437, and an interquartile range of 257–768. Next, we draw five covariates, $x_{1j}-x_{5j}$, collectively $x_j$, which are independently and identically distributed unit normal random variables.

Each aggregate unit is assigned to treatment randomly, using a propensity score that is formed using the first two covariates, with strong separation between treated and comparison units: $\pi(x_j)=\Phi(1-2(x_{1j}>x_{2j})+\zeta_j)$ with $\zeta_j$ a uniform random variable between -.05 and .05. One-third of the sample—1,000 units—are assigned to treatment with probability proportional to $\pi(x_j)$.

Outcomes are composed of five components: $\mu,\tau,u,v,$ and $\varepsilon$. Three of them, $\mu,u,$ and $\varepsilon$, are prognostic (that is, they determine the outcomes for comparison units and are the counterfactual outcomes for treated units) and two of them, $\tau$ and $v$, are treatment effects. Two components, $\mu$ and $\tau$, are covariate-explained, and the other three components, $u,v,$ and $\varepsilon$, are random residual terms, unrelated to any covariates. $\mu$, the prognostic, covariate-explained component, is defined by the nonlinear function $\mu(x_j)=6-12(x_{2j}>0)+|x_{1j}-1|+3x_{5j}$. $\tau$, the heterogeneous, covariate-explained treatment effect, is determined by the function $\tau(x_j)= 1+x_{3j}+.75x_{4j}$. The residual is composed of $u_j \sim N(0,\sigma_u^2)$, the prognostic unit residual; $v_j \sim N(0,\sigma_v^2)$, a unit-specific random treatment effect, and $\varepsilon_j \sim N(0,\frac{\sigma_\varepsilon^2}{w_j})$, the error term, determined by the average of individual-level errors. $v_j$ is included in the DGP for verisimilitude, recognizing that observable covariates alone are unlikely to fully determine treatment effects. 

We then scale each of $\tau,\mu,\varepsilon,u$, and $v$ relative to each other and to a variance decomposition performed on the evaluation data set used for calibration so that results are more easily interpretable. The standard deviation of $y_j$ is targeted to 147, the observed unit-level standard deviation of five-year changes in average Medicare expenditures in our motivating example, measured in dollars per beneficiary per month. The standard deviation of $\tau(x)$ is set to 17, which was the standard deviation of estimated unit-level treatment effects in our motivating example. The standard deviation of $\mu(x)$ is set to 83 to maintain, in combination with $\tau(x)$, a rough $R^2$ of 33 percent, in line with the $R^2$ from our motivating example. We set the default value of $\sigma_u$ to 61 and $\sigma_\varepsilon$ to 2,557 to align with the observed ICC in our motivating example (0.0006), yielding a composite unexplained variance for an average unit that accounts for the remaining 67 percent of the outcome variance. Lastly, we set $\sigma_v$ to 8. This variance component has no analog in our motivating example, so we chose this value as roughly the product of the standard deviations of $u$ and the standard deviation of our prior for treatment effects on the basis that $v$ represents the interaction of treatment status, $z_j$, and $u$, the effect of unobserved or unmeasured covariates on $y$ \citep{si2020}.

Finally, observable outcomes are constructed as: 
\[y_j=\mu(x_j )+z_j \tau(x_j )+u_j+z_j v_j+\varepsilon_j\]
This data-generating process is strongly aligned with the model underlying aBCF; this is an intentional choice because evidence supports that outcomes in the real world can be driven by factors at each level of a hierarchy, including unit-level effects \citep{hedberg2014}, and the purpose of aBCF is to account for these effects. 

For our simulation study, we drew 200 replicate data sets using our DGP and estimated treatment effects for each data set using BCF and aBCF. Both models are given the covariates $x_j$, the true propensity score $\pi_j$ (although in practice usually only an estimated propensity score is available), and the treatment indicator $z_j$. To reduce computation, we do not fit a propensity score model in each replicate data set. This choice is unlikely to influence simulation results because BCF and aBCF use the propensity score identically, so the accuracy of the propensity score should affect both methods equally. aBCF is also given $w_j$ but only as the weighting variable; we did not supply $w_j$ as a covariate to BCF or aBCF. For sensitivity analyses, we draw additional replicate data sets for scenarios in which the prior on $\sigma_u$ is further from the true value or in which the ICC varies in magnitude.

\subsection{Evaluation} \label{subseceval}
We use the simulation results to assess the performance of aBCF along several dimensions. First, we evaluate the accuracy of treatment effect estimates and their standard errors in terms of root mean squared error (RMSE) across simulation replicates, average coverage of a nominal 90 percent interval computed from the 5th and 95th quantiles of the posterior distribution for the relevant estimand across replicates, and average width of the same 90 percent interval across replicates. We evaluate these metrics for two estimands, both of which focus on the set of units receiving the intervention, $T$: unit treatment effects (UTEs) and the sample average treatment effect among the treated (SATT).

UTEs are simply $\tau_j=y_j(1)-y_j(0)$ for each treated unit $j\in T$. We evaluate accuracy of the estimates by taking the RMSE across treated units, calculated as $\sqrt{\frac{\sum_{j \in T}(\hat{\tau}_j-\tau_j)^2}{\sum_{j} z_j=1}}$. We evaluate uncertainty (interval width and coverage) by taking the simple average of each metric (either 90 percent interval width or an indicator for coverage of a 90 percent interval) across treated units within a given simulation replicate. For example, the average UTE 90 percent credible interval width is calculated as $\frac{\sum_{j \in T}width_j^{90}}{\sum_{j} z_j=1}$, where $width_j^{90}$ is the 90 percent credible interval width for unit $j$.

The SATT in each simulation replicate is the average treatment effect for an individual rather than the effect for an aggregate unit. Thus, the SATT is the weighted average of the UTEs, in which the weights are the number of individuals in each aggregate unit, as in the aBCF model: $SATT=\frac{\sum_{j \in T}w_j\hat{\tau}_j}{\sum_{j \in T}w_j}$. This approach is often the most relevant for policy research (for example, researchers are more interested in the effect of an intervention for the average student or Medicare beneficiary rather than the effect for the average school or primary care practice). Even though BCF does not use $w_j$ in model estimation, the weights are still incorporated into the BCF calculation of SATT.

Because in each iteration of the simulation we fit BCF and aBCF to the same data set, we can perform within-iteration comparisons to improve precision of the estimated difference in performance between the two models. We cannot test the RMSE for SATT directly, however, because this approach requires one metric value per data set per model run, and RMSE for SATT is calculated over all replicates. Instead, for the SATT, we test for differences in the squared error, although we still report the RMSE. To evaluate differences in performance, we fit a simple linear model of $m_{ir}=\delta_r+\gamma aBCF_i$, where $m_{ir}$ is the value of the metric $m$ (for example, the interval width) for model run $i$ fit to simulation replicate $r$; $\delta_r$ is a series of fixed effects, one for each simulation replicate; $aBCF_i$ is an indicator for whether model run $i$ used aBCF; and $\gamma$ is the estimate of the difference in performance for aBCF relative to BCF. 

\subsection{Results} \label{subsecresults}
In this section, we describe the performance of aBCF compared with BCF when estimating treatment effects, then describe performance for identifying exemplar units. There is no formal definition for what constitutes an exemplar, and it can be conceptualized in different ways, including units performing above some benchmark (for example, schools whose average test scores improved at least 10 points), or units performing better on a relative scale (for example, the top 25 percent of primary care practices). Lastly, we discuss sensitivity to the prior specification for $\sigma_u$ and evaluate how performance changes with the ICC.
\subsubsection{Treatment effect estimation} \label{subsubsectreat}
For both UTE and SATT estimation, we find that aBCF results in significantly lower RMSE for UTE and lower squared error for SATT (Table~\ref{tabletreatest}). We also find that aBCF achieves significantly narrower uncertainty intervals at the same level of coverage relative to BCF. It is important to note, however, that BCF and aBCF undercover for UTEs, with intervals about 25 percent narrower than they should be. 

Table~\ref{tabletreatest} reports the mean of each metric for each estimand for BCF and aBCF, the results from the fixed effects regression we use to test whether performance differentials are statistically significant, and the percentage improvement for aBCF over BCF. 

\clearpage

\begin{table*}[h]
	\caption{Treatment effect estimation performance}
	\label{tabletreatest}
	\begin{tabular}{@{}cccccc@{}}
		\hline
		Estimand & \multicolumn{1}{c}{Metric}
		& \multicolumn{1}{c}{BCF mean} & \multicolumn{1}{c}{aBCF mean}
		& \multicolumn{1}{c}{Difference} & Percentage \\
		 & & & & (SE) & improvement \\
		\hline
		SATT   & Squared error          & 55.2       & 38.0       & $-$17.2*   & 31\% \\
		       & [RMSE]                 & [7.4]      & [6.2]      & (3.7)      & \\
		SATT   & 90\% interval          & 24.1       & 20.3       & $-$3.8*    & 16\% \\
		       & width                  &            &            & (0.06)     & \\
		SATT   & 90\% interval          & 90.0\%     & 89.0\%     & $-$1\%     & $-$1\% \\
		       & coverage               &            &            & (2.2\%)    & \\
		UTE    & RMSE                   & 17.6       & 16.6       & $-$0.98*   & 6\% \\
		       &                        &            &            & (0.13)     & \\
		UTE    & 90\% interval          & 47.1       & 43.6       & $-$3.5*    & 7\% \\
		       & width                  &            &            & (0.47)     & \\
		UTE    & 90\% interval          & 80.6\%     & 80.4\%     & $-$0.2\%   & 0\% \\
		       & coverage               &            &            & (0.6\%)    & \\
		\hline
	\end{tabular}
	{\raggedright Note: For SATT, we run our performance regression using squared error.
		
		* Indicates the difference between BCF and aBCF is significant at the .05 level in a regression model with simulation replicate fixed effects.
		
		SE = standard error. \par}
\end{table*}

The gains in accuracy and efficiency for SATT that we see for aBCF in Table~\ref{tabletreatest}, relative to BCF, derive mostly from the heteroskedastic variance term. The addition of unit random effects has a relatively small bearing on the SATT for two reasons: they affect counterfactual and treated outcomes equally, and they necessarily have and expected mean of zero across the entire sample. Nonetheless, including the weights improves estimation by allowing the model to better fit the data, and the random effects improve the modeling of the prognostic portion of outcomes.

\subsubsection{Identification of units with exemplary outcomes} \label{subsubsecexemplar}
The unit random effects themselves do not contribute to the estimated treatment effects, but they quantify a key aspect of performance that is often relevant to policymakers: unit-specific unexplained performance on outcomes of interest. aBCF targets these applications by including the $u_j$, which separate individual-level error, a form of noise, from the unexplained performance of the aggregate unit. 

In our simulation, we define exemplars according to the sum of $u_j$ and $v_j$, incorporating unexplained unit-specific performance prognostically and in treatment effects. With aBCF, we identify top performers based on $\hat{u}_j$ (that is, based on aBCF’s estimate of unit effects only; because aBCF cannot distinguish between unit prognostic and unit treatment effects, $\hat{u}_j$ corresponds to a combination of $u_j$ and $v_j$ in the DGP). With BCF, we use the overall residual $\hat{\varepsilon}_j$, which, because of the mismatch between DGP and model because BCF does not have a mechanism for disentangling these aspects of performance that are not explained by covariates, includes both unit effects and noise $u_j$, $v_j$, and $\varepsilon_j$ in the DGP). It is therefore unsurprising that aBCF substantially outperforms BCF at identifying top-performing units by this metric. Neither model is fully able to target the correct estimand because neither account for unit-specific random treatment effects $v_j$.

\begin{figure}[!h]
	\caption{Estimation of residuals}
	\includegraphics[width=.9\textwidth]{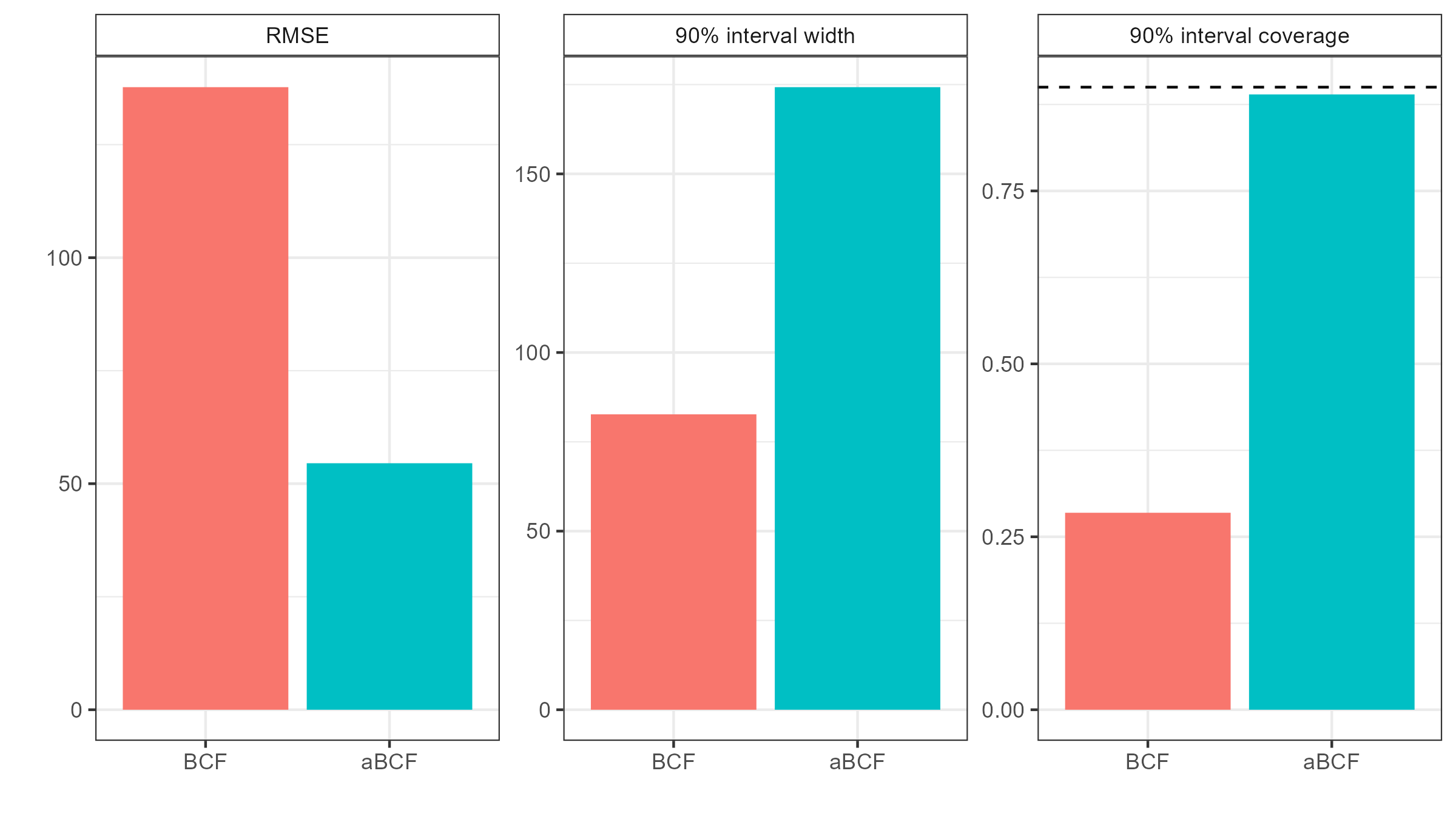}
	\label{figresid}
\end{figure}

In Figure~\ref{figresid}, we show RMSE, interval width, and coverage for the estimates of $u_j+v_j$ derived from BCF and aBCF. One might justly observe that the comparison is unfair to BCF because the data-generating process, which includes $u_j$ terms, clearly favors aBCF. To the extent that the data-generating process’s assumption is realistic—as indeed we believe it is realistic to assume that counterfactual outcomes vary beyond what covariates explain—the comparison illustrates BCF’s unsuitability for this application. This shortcoming was a primary motivation for the development of aBCF. 

As in estimation of SATT and UTE, aBCF significantly outperforms BCF on RMSE, although much more substantially because this is where the model changes are most salient since they directly affect identification of residuals. Unlike for treatment effect estimation, here aBCF does show large improvements in coverage, although these come mostly from appropriately expanding uncertainty intervals.

To further assess performance, we also explore the identification of the specific highest-performing units as measured by the rank of $u_j+v_j$. In addition to generally assessing units’ specific performance, identifying top performers can be relevant for policy evaluations, in which researchers might wish to select units for further study or follow-up using a qualitative approach to find out what drives quantitatively unexplained performance. In this analysis, we find that among the units aBCF ranks in the top quartile using $\hat{u}_j$, 63 percent are actually in the top quartile (based on true values of $u_j+v_j$), compared with 45 percent of those that BCF ranks based on $\hat{\varepsilon}_j$, which is a substantial improvement.

\subsubsection{Prior sensitivity} \label{subsubsecpriorsens}
Of the two extensions that distinguish aBCF from BCF—heteroskedastic error and unit random effects—only the unit random effects require us to add new prior distributions. In this section, we explore aBCF’s sensitivity to the hyperprior for the standard deviation of these random effects, $\sigma_u$. By default, we use a prior that is half-normal with scale $\frac{2sd(y)}{3}$, where $y$ is the outcome variable, and the standard deviation calculation is weighted using $w_j$. In the updated $abcf$ R package, we allow the user to override this default because we believe it should also be possible to obtain a reasonable prior for $\sigma_u$ by obtaining information about the likely unit-level variance from similar data sets, if they are available.

Because we chose this default distribution and calibrated the simulation DGP based on the same evaluation data set, the prior and simulation are well aligned: the true value of $\sigma_u$ is 61, the mean of the prior is 78, and the median is 66. Real-world applications are unlikely to show such fortuitous alignment, so we test sensitivity to this prior by repeating the simulation using an array of priors, with a scale ranging from one-quarter to four times as large as the default (that is, from $\frac{sd(y)}{6}$ to $\frac{8sd(y)}{3}$. 

When comparing across prior specifications, we find that there are no differences in performance on RMSE, coverage, or interval width for UTE or SATT. For estimation of $\sigma_u$ itself, we find that coverage and RMSE for $\sigma_u$ are statistically significantly worse when the prior is set to one-quarter of the default scale, although these differences are not strikingly large because of the scale of the data (see Table~\ref{tablepriorsens}). Varying the hyperprior specifications for $\sigma_u$ does not affect estimation of $\sigma_\varepsilon$ (not shown).

\begin{table*}[h]
	\caption{Parameter estimation under different $\sigma_u$ priors}
	\label{tablepriorsens}
	\begin{tabular}{@{}cccc@{}}
		\hline
		Hyperprior scale & 
		\multicolumn{1}{c}{RMSE of $\sigma_u$} &
		\multicolumn{1}{c}{Interval width for $\sigma_u$} & 
		\multicolumn{1}{c}{Coverage for $\sigma_u$} \\
		\hline
		$\frac{1}{4}x$   & 6.11*          & 16.6*          & 82\%    \\
        $\frac{1}{2}x$   & 5.23           & 16.5           & 89.5\%  \\
        \textit{Default} & \textit{5.13}  & \textit{16.4}  & \textit{91\%}    \\
        $2x$             & 5.06           & 16.4           & 88\%    \\
        $4x$             & 5.06           & 16.4           & 90\%    \\
		\hline
	\end{tabular}
	
	{\raggedright * This indicates the difference from the default is significant at the .05 level in a regression model with simulation replicate fixed effects. For RMSE, significance testing is performed on the squared error values from each model run. \par}
\end{table*}

We are encouraged that $\sigma_u$ is fairly well identified by the data using real-world variability in aggregate unit sample sizes and that, as a result, aBCF is not particularly sensitive to the specification of this prior.

\subsubsection{Sensitivity to ICC} \label{subsubsecicc}
One of the goals of aBCF is to account for clustering of individuals within aggregate units. The strength of clustering in the data might affect the performance of aBCF relative to BCF. This section assesses the importance of this factor. When the ICC is 0 (that is, $\sigma_u=0$), all individuals within a unit are entirely independent, and heteroskedasticity is most pronounced. In this case, we expect BCF fit with aggregate data to underperform, and BCF fit with individual data would be correct, an alternative we did not explore for computational reasons because computational intractability of these types of analyses using individual-level data is a key motivating factor for developing aBCF. When the ICC is 1 (that is, $\sigma_\varepsilon=0$), BCF fit to aggregate data is correct because there is no heteroskedasticity, and each unit’s residual variance is just $\sigma_u^2$. We expect that BCF will perform best when the ICC is 1 and worst when the ICC is 0. Because aBCF is flexible enough to represent all values of the ICC simply by varying the estimates of $\sigma_u$ and $\sigma_\varepsilon$, we expect it to perform well across the board, except perhaps in cases where the prior is much smaller than the true value, per the prior sensitivity results in Section~\ref{subsubsecpriorsens}.

Although ICC is a natural way of conceptualizing the degree of clustering at the individual level, when working at the aggregate level, we find it easier to think in terms of the share of the aggregate residual variance that is explained by average individual error $\varepsilon_j$ versus the aggregate residual $u_j$. This proportion is determined by the ICC and the average sample size. For example, in our main DGP, the ICC is only 0.0006 ($\sigma_u=61; \sigma_\varepsilon=2557$), but, given the large sample sizes (average number of individuals per unit $\bar{w} = 608$), on average, 26 percent of a unit’s residual variance is attributable to the unit random effects. For this sensitivity analysis, we check performance across a range of residual variance shares ranging from 0 to 100 percent (and corresponding ICCs of 0 to 1). 

\begin{figure}[H]
	\caption{Estimation for varying residual variance shares}
	\includegraphics[width=.9\textwidth]{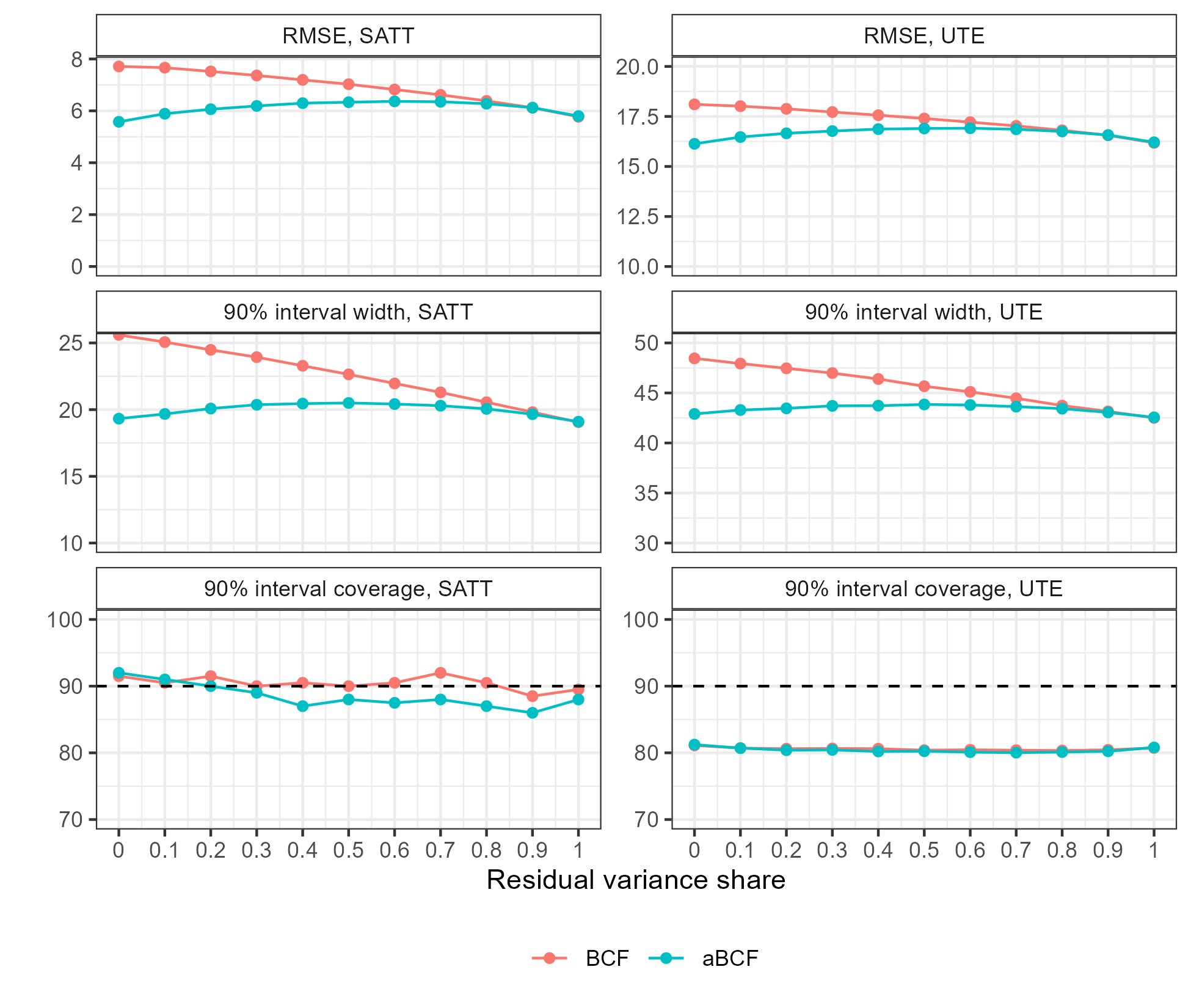}
	\label{figicc}
\end{figure}

As Figure~\ref{figicc} shows, aBCF’s benefits in the aggregate data setting are highest when the residual variance share is lowest and heteroskedasticity is most pronounced. There are no differences in coverage for SATT or UTEs, but the differences in squared error and interval width are significant for residual variance shares from 0 to 70 percent, except for SATT interval width, which is significantly lower for aBCF through 90 percent residual variance share. 

Importantly, we do not see evidence that aBCF underperforms compared with BCF when the residual variance share is one (ICC = 1). The persistent strength of aBCF is not surprising because aBCF can collapse down to the BCF specification by estimating $\sigma_\varepsilon=0$, but it is reassuring to see that, in practice, the model correctly allocates variance between the two available components.

\section{Discussion} \label{secdiscuss}
Hierarchical data are ubiquitous in policy evaluations, which can study everything from patients nested within health care practices or students nested within schools to employees nested within organizations or residents nested within states. For computational reasons, working directly with individual-level data (that is, patients, students, employees, or residents) is often prohibitive. Aggregate data (that is, health care practices, schools, organizations, states) offer distinct computational advantages, as well as theoretical ones from better aligning analysis to the level of selection (although accounting for aggregate selection in individual models is possible). aBCF provides a powerful new method for working with aggregated data that flexibly adjusts for confounding relationships, detects and estimates heterogeneity, and accounts appropriately for heteroskedasticity. aBCF’s handling of heteroskedastic errors improves estimation of SATTs, lowering RMSE and interval widths while maintaining coverage, as well as lowering RMSE and interval widths for UTEs. In addition, aBCF’s unit random effects enable identification of top-performing units on the basis of their unexplained performance, allowing researchers to further explore unmeasured outcome relationships. Our simulation study shows that aBCF is robust to different prior assumptions and degrees of clustering.

Although our simulation study shows that aBCF improves on BCF’s ability to estimate UTEs, it falls short of fully capturing these effects. Specifically, aBCF’s estimates of unit random effects ($\hat{u}_j$) do not distinguish whether a unit performed exceptionally well for reasons unrelated to, or in response to, the intervention (that is, $u_j$ versus $v_j$ in the data-generating process). In separate simulations (discussed in appendix \ref{apxute}) we propose a model that makes this distinction, estimating unit random effects and unit random treatment effects, with some assumed (positive or negative) correlation between them. These additional parameters—the unit random treatment effects and correlation coefficient—were not identifiable by the data in the settings we considered, resulting in considerable sensitivity to the priors used. Appendix \ref{apxute} describes our approach to estimating these terms and we provide code for their estimation in the $abcf$ package. Estimation of unit-specific treatment effects is the next frontier for heterogeneous treatment effect estimation and could lead to substantial improvements in coverage for UTEs in future work. 

Lastly, although aBCF incorporates one batch of random effects, effectively allowing for correlations of residuals for individuals within an aggregate unit, we believe there is value in allowing for correlation across other dimensions and more complex hierarchical structures. For example, such extensions could account for repeated measures (by including unit-time random effects), which would open the door to true longitudinal analysis or for more complicated nesting, such as schools within school districts or counties within states. aBCF represents a valuable tool in its own right, and a stepping stone toward ever-advancing techniques for estimating treatment effect heterogeneity.

\bibliographystyle{ba}
\bibliography{aBCF}

\begin{thebibliography}{20}
\newcommand{\enquote}[1]{``#1''}
\expandafter\ifx\csname natexlab\endcsname\relax\def\natexlab#1{#1}\fi
\expandafter\ifx\csname url\endcsname\relax
  \def\url#1{{\tt #1}}\fi
\expandafter\ifx\csname urlprefix\endcsname\relax\def\urlprefix{URL }\fi
\ifx\endbibitem\undefined \let\endbibitem\relax\fi

\bibitem[{Abadie et~al.(2017)Abadie, Athey, Imbens, and
  Wooldridge}]{abadie2017}
Abadie, A., Athey, S., Imbens, G., and Wooldridge, J. (2017).
\newblock \enquote{When Should You Adjust Standard Errors for Clustering?}
\newblock {\em National Bureau of Economic Research Working Paper\/}, 24003.
\endbibitem

\bibitem[{Bargagli-Stoffi et~al.(2023)Bargagli-Stoffi, Cadei, Lee, and
  Dominici}]{bargagli2023}
Bargagli-Stoffi, F., Cadei, R., Lee, K., and Dominici, F. (2023).
\newblock \enquote{Causal Rule Ensemble: Interpretable Discovery and Inference
  of Heterogenous Treatment Effects.}
\newblock {\em https://doi.org/10.48550/arXiv.2009.09036\/}.
\endbibitem

\bibitem[{Chetty et~al.(2014)Chetty, Friedman, and Rockoff}]{chetty2014}
Chetty, R., Friedman, J., and Rockoff, J. (2014).
\newblock \enquote{Measuring the Impacts of Teachers II: Teacher Value-Added
  and Student Outcomes in Adulthood.}
\newblock {\em American Economic Review\/}, 104(9): 2633--2679.
\endbibitem

\bibitem[{Chipman et~al.(1998)Chipman, George, and McCulloch}]{chipman1998}
Chipman, H., George, E., and McCulloch, R. (1998).
\newblock \enquote{Bayesian CART Model Search.}
\newblock {\em Journal of the American Statistical Association\/}, 93(443):
  935--948.
\endbibitem

\bibitem[{Chipman et~al.(2010)Chipman, George, and McCulloch}]{chipman2010}
--- (2010).
\newblock \enquote{BART: Bayesian Additive Regression Trees.}
\newblock {\em Annals of Applied Statistics\/}, 4(1): 266--298.
\endbibitem

\bibitem[{Deke(2016)}]{deke2016}
Deke, J. (2016).
\newblock \enquote{Design and Analysis Considerations for Cluster Randomized
  Trials That Have a Small Number of Clusters.}
\newblock {\em Evaluation Review\/}, 40(5): 444--486.
\endbibitem

\bibitem[{Dorie et~al.(2019)Dorie, Hill, Shalit, Scott, and
  Cervone}]{dorie2019}
Dorie, V., Hill, J., Shalit, U., Scott, M., and Cervone, D. (2019).
\newblock \enquote{Automated versus Do-It-Yourself Methods for Causal
  Inference: Lessons Learned from a Data Analysis Competition.}
\newblock {\em Statistical Science\/}, 34(1): 43--68.
\endbibitem

\bibitem[{Gelman et~al.(1996)Gelman, Roberts, and Gilks}]{gelman1996}
Gelman, A., Roberts, G., and Gilks, W. (1996).
\newblock \enquote{Efficient Metropolis Jumping Rules.}
\newblock {\em Bayesian Statistics\/}, 5: 599--607.
\endbibitem

\bibitem[{Hahn et~al.(2020)Hahn, Murray, and Carvalho}]{hahn2020}
Hahn, P., Murray, J., and Carvalho, C. (2020).
\newblock \enquote{Bayesian Regression Tree Models for Causal Inference:
  Regularization, Confounding, and Heterogeneous Effects.}
\newblock {\em Bayesian Analysis\/}, 15(3): 965--1056.
\endbibitem

\bibitem[{Hedberg and Hedges(2014)}]{hedberg2014}
Hedberg, E. and Hedges, L. (2014).
\newblock \enquote{Reference Values of Within-district Intraclass Correlations
  of Academic Achievement by District Characteristics: Results from a
  Meta-Analysis of District-Specific Values.}
\newblock {\em Evaluation Review\/}, 38(6): 546--582.
\endbibitem

\bibitem[{Hill(2011)}]{hill2011}
Hill, J. (2011).
\newblock \enquote{Bayesian Nonparametric Modeling for Causal Inference.}
\newblock {\em Journal of Computational and Graphical Statistics\/}, 20(1):
  217--240.
\endbibitem

\bibitem[{Hill et~al.(2020)Hill, Linero, and Murray}]{hill2020}
Hill, J., Linero, A., and Murray, J. (2020).
\newblock \enquote{Bayesian Additive Regression Trees: A Review and Look
  Forward.}
\newblock {\em Annual Review of Statistics and Its Application\/}, 7(1):
  251--278.
\endbibitem

\bibitem[{Imbens and Rubin(2015)}]{imbens2015}
Imbens, G. and Rubin, D. (2015).
\newblock {\em Causal Inference in Statistics, Social, and Biomedical
  Sciences\/}.
\newblock Cambridge, UK: Cambridge University Press.
\endbibitem

\bibitem[{Kane and Staiger(2008)}]{kane2008}
Kane, T. and Staiger, D. (2008).
\newblock \enquote{Estimating Teacher Impacts on Student Achievement: An
  Experimental Evaluation.}
\newblock {\em National Bureau of Economic Research Working Paper\/}, 14607.
\endbibitem

\bibitem[{Killip et~al.(2004)Killip, Mahfoud, and Pearce}]{killip2004}
Killip, S., Mahfoud, Z., and Pearce, K. (2004).
\newblock \enquote{What Is an Intracluster Correlation Coefficient? Crucial
  Concepts for Primary Care Researchers.}
\newblock {\em Annals of Family Medicine\/}, 2(3): 204--208.
\endbibitem

\bibitem[{Si et~al.(2020)Si, Trangucci, Gabry, and Gelman}]{si2020}
Si, Y., Trangucci, R., Gabry, J., and Gelman, A. (2020).
\newblock \enquote{Bayesian Hierarchical Weighting Adjustment and Survey
  Inference.}
\newblock {\em Survey Methodology\/}, 46(2): 181--241.
\endbibitem

\bibitem[{{Stan Development Team}(2020)}]{stan2020}
{Stan Development Team} (2020).
\newblock \enquote{Prior Choice Recommendations.}
\newblock {\em Stan wiki
  \url{https://github.com/stan-dev/stan/wiki/Prior-Choice-Recommendations}\/}.
\endbibitem

\bibitem[{Swankoski et~al.(2022)Swankoski, O'Malley, Tu, Petersen, Singh,
  Geonnotti, Keith, Dale, Morrison, Peikes, McCall, Duda, Markovitz, Heitkamp,
  Lee-Morrison, Peebles, Sarwar, Shin, and Fu}]{swankoski2022}
Swankoski, K., O'Malley, A., Tu, H., Petersen, D., Singh, P., Geonnotti, K.,
  Keith, R., Dale, S., Morrison, N., Peikes, D., McCall, N., Duda, N.,
  Markovitz, A., Heitkamp, S., Lee-Morrison, K., Peebles, V., Sarwar, R., Shin,
  E., and Fu, N. e.~a. (2022).
\newblock \enquote{Independent Evaluation of Comprehensive Primary Care Plus
  (CPC+) Fourth Annual Report.}
\newblock {\em U.S. Department of Health and Human Services
  \url{https://www.cms.gov/priorities/innovation/data-and-reports/2022/cpc-plus-fourth-annual-eval-report}\/}.
\endbibitem

\bibitem[{Thal and Finucane(2023)}]{thal2023}
Thal, D. and Finucane, M. (2023).
\newblock \enquote{Causal Methods Madness: Lessons Learned from the 2022 ACIC
  Competition to Estimate Health Policy Impacts.}
\newblock {\em Observational Studies\/}, 9(3): 3--27.
\endbibitem

\bibitem[{Yeager et~al.(2019)Yeager, Hanselman, Walton, Murray, Crosnoe,
  Muller, Tipton, Schneider, Hulleman, Hinojosa, Paunesku, Romero, Flint,
  Roberts, Trott, Iachan, Buontempo, Man~Yang, and Carvalho}]{yeager2019}
Yeager, D., Hanselman, P., Walton, G., Murray, J., Crosnoe, R., Muller, C.,
  Tipton, E., Schneider, B., Hulleman, C., Hinojosa, C., Paunesku, D., Romero,
  C., Flint, K., Roberts, A., Trott, J., Iachan, R., Buontempo, J., Man~Yang,
  S., and Carvalho, C. e.~a. (2019).
\newblock \enquote{A National Experiment Reveals Where a Growth Mindset
  Improves Achievement.}
\newblock {\em Nature\/}, 573: 364--369.
\endbibitem

\end{thebibliography}

\appendix
\section{Unit random treatment effects} \label{apxute}
As section \ref{subsubsectreat} shows, BCF and aBCF substantially undercover the UTEs in our simulation study. This phenomenon has also been observed in other settings, including results from the 2017 American Causal Inference Conference challenge \citep{hahn2020} and 2022 American Causal Inference Conference challenge \citep{thal2023}. Overstating statistical precision undermines the policy goals UTE estimation seeks to achieve. For example, UTEs can identify units that benefit the most from interventions. In one application, policymakers can use this information to predict the effect of selectively expanding an intervention to a target population most likely to benefit from it. In another, policymakers might seek to reward individual units for improving in response to an intervention, as the Centers for Medicare \& Medicaid Services has done under several recent initiatives (for example, Comprehensive Primary Care Plus and Primary Care First). Without an accurate understanding of the uncertainty of these UTEs, policymakers could ground policy decisions in overconfident results that ultimately waste public funds.

We posit unit random treatment effects as one explanation for the consistent undercoverage of UTEs. Such effects could arise because of true random idiosyncrasy or effect modification from unobserved covariates (for example, a school principal or a health care practice manager who really believes in the intervention). Unit random treatment effects extend the idea that unobserved factors and idiosyncrasy, encapsulated in the $u_j$ term in the aBCF specification, can drive counterfactual outcomes. Indeed, it seems naïve to assume that unobserved factors only affect prognostic levels of outcomes and not treatment effects (in other words, to believe that treatment effects are wholly predictable from observable covariates).

To align our study with our intuition that covariates are unlikely to fully explain treatment effects, we added unit random treatment effects to our simulation DGP to represent the portion of each unit’s response to treatment not captured by observable covariates, which we call $v_j \sim N(0,\sigma_v^2)$. We incorporated these effects into the DGP (described in section \ref{subsecdgp}) that we used to evaluate aBCF, multiplied by the treatment indicator $z_j$. Under this DGP, aBCF fell short on coverage of UTEs (see Table \ref{tabletreatest}), as we would expect from a specification that does not attempt to model unit random treatment effects. To achieve correct coverage, we attempted to adapt aBCF to account for these unit random treatment effects. We call this adapted model individualized BCF (iBCF) because it provides treatment effect estimates for each individual unit. iBCF makes several changes to the aBCF model.

First, we add two parameters to our model: $v_j$, the random treatment effect for each practice, and $\sigma_v$, the scale of these effects:
\[ y_j=\mu(x_j,\hat{\pi}_j)+z_j \tau(x_j )+u_j+z_j v_j+\varepsilon_j;\]
\[ \varepsilon_j \sim N(0,\frac{\sigma_\varepsilon^2}{w_j})\]
\[ u_j \sim N(0,\sigma_u^2 )\]
\[ v_j \sim N(0,\sigma_v^2)\]

This change by itself implies that the total residual variance in the treatment group ($\frac{\sigma_\varepsilon^2}{w_j} +\sigma_u^2+\sigma_v^2$) can never be smaller than the residual variance in the comparison group ($\frac{\sigma_\varepsilon^2}{w_j} +\sigma_u^2$). This assumption is unrealistically strict, so we posit a correlation $\rho$ between the two random effects $u$ and $v$, which allows the residual variance in the treatment group ($\frac{\sigma_\varepsilon^2}{w_j} +\sigma_u^2+\sigma_v^2+2\rho\sigma_u \sigma_v$) to be lower or higher than it is in the comparison group, depending on the value of $\rho$. Thus, the unit random effects are modeled as:

\[
\begin{bmatrix}
	u_j \\ v_j
\end{bmatrix} 
\sim N(
\begin{bmatrix}
	0 \\ 0
\end{bmatrix} 
,
\begin{bmatrix}
	\sigma_u^2 & \rho\sigma_u\sigma_v \\
	\rho\sigma_u\sigma_v & \sigma_v^2 
\end{bmatrix} 
)\]
For example, $u$ and $v$ could be positively correlated if primary care practices with higher-quality care coordinators (an unmeasured factor) have better outcomes in general because of those care coordinators (high $u_j$) and can leverage the coordinators to better implement the intervention, thus deriving more benefit from it (high $v_j$). By contrast, $u$ and $v$ could be negatively correlated if the intervention trained care coordinators to meet some minimum standard. Here, practices that already had good care coordinators (and hence a high $u_j$) would benefit the least from treatment (low $v_j$), and practices that had worse care coordinators would benefit the most.

To incorporate these new terms into the model, we specify prior distributions for $\sigma_v$ and $\rho$. For $\rho$, we specify a weakly informative recentered beta prior \citep{stan2020}: $\rho \sim 2B(2,2)-1$. Because, to our knowledge, these random treatment effects have never been studied before, there is no literature to provide an evidence-based prior for $\sigma_v$. Instead, we use the structured prior logic of \citet{si2020} to conceptualize $v_j$ as an interaction effect between general idiosyncrasies, $u_j$, and treatment, $z_j$. \citet{si2020} propose that a reasonable prior for the scale of interaction effects can be formed using the product of the scales of the two parent main effects. For example, if we are interested in interactions between state s and race r, we can use as a prior for $\sigma_{s\times r} \sim N^+ (0,\lambda_s^2 \lambda_r^2 \sigma^2)$, where $\lambda_s$ is the unitless scale of state effects (for example, taking the standard deviation of state effects and dividing by the standard deviation of the outcome, $\frac{\sigma_s}{sd(y)}$, $\lambda_r$ is the unitless scale of race effects, and $\sigma$ is a general scale term, to bring us back to natural units. Under this formulation, our priors for the interactions update as the model learns from the data about the importance of the main effects; if there is evidence of large state effects, then larger state-race interactions are more plausible, but if both state and race effects are infinitesimal, it is highly unlikely that their interactions are hugely important.

We adapt this logic to our case using a half-normal prior whose scale is the product of the scale of unit random effects $\sigma_u$, and the scale of average treatment effects across a superpopulation of interventions, $\psi$: $\sigma_v \sim N^+ (0,\frac{\sigma_u^2}{Var(y)} \frac{\psi^2}{Var(y)} Var(y))$. Thus, the prior for $\sigma_v$ updates as the iBCF model runs as we learn from the model how important unit-specific effects are in general via $\sigma_u$. Note that $\psi$ is not the range of treatment effects estimated by the model—in other words, $sd(\tau(x))$. Rather, it is the standard deviation of the superpopulation of SATTs across interventions (that is, the standard deviation of the prior for the SATT). Because the model is fit to only one intervention, we cannot estimate $\psi$ from the data. Instead, we must provide an estimate of $\psi$ as part of our prior specification, ideally derived from an external meta-analysis. In our DGP, which is based on Medicare primary care model evaluations, we use $\psi=25$ based on the authors’ unpublished meta-analysis of prior literature. In our simulation, the value of $\sigma_v$ is based on this same logic (we use $\sigma_v=8$ on the basis of that being about the mean of $N^+ (0,\frac{61^2}{147^2}\frac{25^2}{147^2}147^2$), so our prior and the DGP match exactly. 

\clearpage
\begin{table*}[!h]
	\caption{iBCF estimation performance}
	\label{tableibcfperf}
	\begin{tabular}{@{}cccccc@{}}
		\hline
		Estimand & \multicolumn{1}{c}{Metric}
		& \multicolumn{1}{c}{aBCF mean} & \multicolumn{1}{c}{iBCF mean}
		& \multicolumn{1}{c}{Difference} & Percentage \\
		& & & & (SE) & improvement \\
		\hline
		SATT   & Squared error          & 38.0       & 37.8       & $-$0.21    & 0\% \\
		&                        &            &            & (1.8)      & \\
		SATT   & 90\% interval          & 20.3       & 20.3       & $-$0.01    & 0\% \\
		& width                  &            &            & (0.03)     & \\
		SATT   & 90\% interval          & 89.0\%     & 89.0\%     & 0          & 0\% \\
		& coverage               &            &            & (0.01\%)   & \\
		UTE    & RMSE                   & 16.6       & 16.6       & 0.002      & 0\% \\
		&                        &            &            & (0.013)    & \\
		UTE    & 90\% interval          & 43.6       & 53.8       & 10.3*      & 24\% \\
		& width                  &            &            & (0.20)     & \\
		UTE    & 90\% interval          & 80.4\%     & 89.1\%     & 8.7\%      & 11\% \\
		& coverage               &            &            & (0.3\%)    & \\
		\hline
	\end{tabular}
	{\raggedright Note: For SATT, we run our performance regression using squared error.
		
		* Indicates the difference between aBCF and iBCF is significant at the .05 level in a regression model with simulation replicate fixed effects.
		
		For iBCF, the prior for $\sigma_v$ is perfectly calibrated to the value used in the DGP. \par}
\end{table*}

Table~\ref{tableibcfperf} compares the performance on SATT and UTE estimation between iBCF and aBCF. With a perfectly calibrated prior distribution, our simulation study showed that, compared with aBCF, iBCF significantly improves coverage of the UTEs, bringing them up to nominal coverage. We do not see any change to RMSE of the UTEs, however, meaning that the coverage improvements come solely from achieving better-calibrated (and therefore much wider) uncertainty intervals. Because the estimates of the UTEs are not any closer to the true values, this implies that we are not actually able to estimate any of the individual $v_j$, so we cannot discern which units have high or low idiosyncratic responses to treatment. Still, the ability to achieve proper uncertainty calibration for UTEs would be a tremendous advance.

\clearpage
\begin{figure}[!h]
	\caption{iBCF variance parameter estimation}
	\includegraphics[width=.9\textwidth]{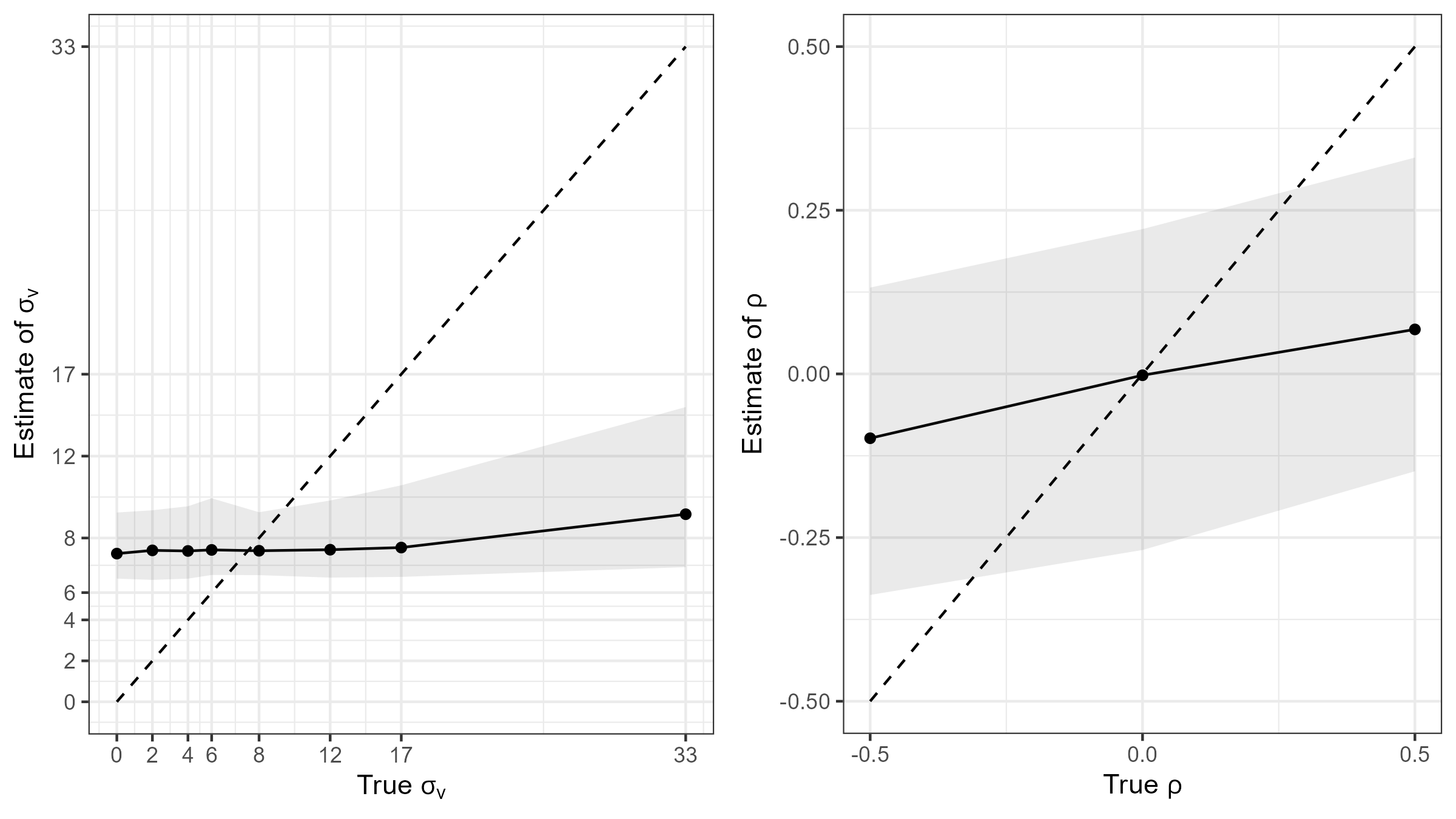}
	\label{figsigv}
\end{figure}

Results with imperfectly calibrated prior distributions were less encouraging. We ran versions of the simulation that varied the true values of $\sigma_v$ and $\rho$, leaving the priors fixed as described above, to determine how well we can estimate these quantities from the data. Figure~\ref{figsigv} compares the true and estimated values of $\sigma_v$ and $\rho$ in these scenarios. It is clear from the figure that the model is unable to learn the correct distributions from the data, instead returning the values from the prior. Thus, the $v_j$ and $\rho$ are not identified. In the left panel, iBCF is almost entirely insensitive to the true value of $\sigma_v$, consistently estimating values around 8—the prior mean—when the truth is anywhere from 0 to 17, with only a barely perceptible change in estimates when the true value is 33. In the right panel, we see that when the true value of $\rho$ is -0.5 or 0.5, the model also barely responds, returning a mean estimate of only -0.1 or 0.1, respectively. Because of the relative weakness of the $2B(2,2)-1$ prior, the estimate’s insensitivity to different scenarios indicates a lack of identification in the data instead of an overly tight prior. 

Without the ability to identify $\sigma_v$ or $\rho$ from the data, we are entirely reliant on the priors. The structured prior formulation in \citet{si2020} is a principled approach; we can ground our value of $\psi$ based on a meta-analysis, and, as section \ref{subsubsecpriorsens} shows, we can estimate $\sigma_u$ with reasonable accuracy. Still, estimating random treatment effects solely on the basis of the prior is unappealing in applied practice.

We are optimistic that our exploration of unit random treatment effects will inform future work on this important problem. Properly calibrated uncertainty for UTEs would open the door to better understanding of intervention scaling effects, more accurate treatment response rewards, and better understanding of effects for more granular subgroups created from post-hoc fitting of UTEs (for example, \citealp{bargagli2023}).

\section{Markov Chain Monte Carlo sampler} \label{apxmcmc}
The modifications to BCF for both aBCF and iBCF require some reworking of the Markov Chain Monte Carlo (MCMC) sampler. For BCF, the model can be written as: 
\[y_j \sim N(\mu(x_j,\hat{\pi}_j)+z_j \tau(x_j), \sigma_\varepsilon^2)\]
For aBCF, that becomes:
\[y_j \sim N(\mu(x_j,\hat{\pi}_j)+z_j \tau(x_j), \frac{\sigma_\varepsilon^2}{w_j} +\sigma_u^2)\]
And for iBCF, it becomes:
\[y_j \sim N(\mu(x_j,\hat{\pi}_j)+z_j \tau(x_j), \frac{\sigma_\varepsilon^2}{w_j} +\sigma_u^2+z_j \sigma_v^2+z_j 2\rho\sigma_u \sigma_v)\]
For simplicity, we define $f(x_j)= \mu(x_j,\hat{\pi}_j)+z_j \tau(x_j)$. Additionally, we hereafter refer to the new compound variance as $\sigma_j^2$, with $\sigma_j^2=  \frac{\sigma_\varepsilon^2}{w_j} +\sigma_u^2$ for aBCF and $\sigma_j^2=\frac{\sigma_\varepsilon^2}{w_j} +\sigma_u^2+z_j \sigma_v^2+z_j 2\rho\sigma_u \sigma_v$ for iBCF. 

Because of the compound variance, the update for $\sigma_\varepsilon$ is no longer conjugate, nor are the updates for $\sigma_u$, $\sigma_v$, or $\rho$, so we use an adaptive Metropolis-in-Gibbs update for each parameter. We target an acceptance rate of 44 percent because this is optimal for single parameters \citep{gelman1996}, and we recalculate the standard deviation of the proposals every 100 MCMC iterations. In addition, this marginalized specification allows us to make simpler draws of $u$ and $v$ after fitting the rest of the model, without needing to calculate a posterior conditional that incorporates $u$ and $v$.

\subsection{Posterior conditional for $\sigma_\varepsilon$}
The prior for $\sigma_\varepsilon$ remains
\[\sigma_\varepsilon^2 \sim InverseGamma(\frac{\nu}{2},\frac{\nu\lambda}{2})\]

as it is in BCF; the changes described in section \ref{subsecmodel} alter the way $\nu$ and $\lambda$ are calibrated, but not the prior itself. The likelihood is:
\[\prod_{j=1}^{n} N(y_j |f(x_j),\sigma_j^2 )\]

Yielding a conditional posterior of:
\[Pr(\sigma_\varepsilon^2|\dots) \propto(\sigma_\varepsilon^2)^{(-\frac{\nu}{2}-1)}exp(-\frac{\nu \lambda}{2\sigma_\varepsilon^2})\prod_{j=1}^{n}(\sigma_j^2)^{-\frac{1}{2}}exp(-\frac{1}{2}\sum_{j=1}^{n}\frac{(y_j-f(x_j))^2}{\sigma_j^2})\]

\subsection{Posterior conditional for $\sigma_u$}
We chose a half-normal prior for $\sigma_u$ with a scale $s_u$. As discussed in the section \ref{subsubsecpriorsens}, we chose $s_u=\frac{2sd(y)}{3}$ as a default. Thus, the prior is:
\[\sigma_u \sim N^+ (0,s_u^2)\]

The likelihood is the same as for $\sigma_\varepsilon$:
\[\prod_{j=1}^{n} N(y_j |f(x_j),\sigma_j^2 )\]

Yielding a conditional posterior of:
\[Pr(\sigma_u^2|\dots) \propto exp(-\frac{\sigma_u^2}{2s_u^2})\prod_{j=1}^{n}(\sigma_j^2)^{-\frac{1}{2}}exp(-\frac{1}{2}\sum_{j=1}^{n}\frac{(y_j-f(x_j))^2}{\sigma_j^2})\]

\subsection{Posterior conditional for $\sigma_v$}
As discussed in Appendix~\ref{apxute}, we use a structured prior for $\sigma_v$ following \citet{si2020}, where the scale $s_v$ of the half-normal prior is determined by the scale of unit random effects $\sigma_u$ and the prior scale of average treatment effects across a superpopulation of interventions, $\psi$:
\[s_v=\frac{\sigma_u}{sd(y)}   \frac{\psi}{sd(y)}  sd(y)\]

Thus, the prior is:
\[\sigma_v \sim N^+ (0,s_v^2)\]

The likelihood is the same as for $\sigma_\varepsilon$ and $\sigma_u$:
\[\prod_{j=1}^{n} N(y_j |f(x_j),\sigma_j^2 )\]

Yielding a conditional posterior of:
\[Pr(\sigma_v^2|\dots) \propto exp(-\frac{\sigma_v^2}{2s_v^2})\prod_{j=1}^{n}(\sigma_j^2)^{-\frac{1}{2}}exp(-\frac{1}{2}\sum_{j=1}^{n}\frac{(y_j-f(x_j))^2}{\sigma_j^2})\]

\subsection{Posterior conditional for $\rho$}
We use a rescaled beta prior for $\rho$:
\[\rho \sim 2B(2,2)-1\]

The likelihood is the same as for $\sigma_\varepsilon$, $\sigma_u$, and $\sigma_v$:
\[\prod_{j=1}^{n} N(y_j |f(x_j),\sigma_j^2 )\]
Yielding a conditional posterior of:
\[Pr(\rho|\dots) \propto (\rho+1)(\rho-1)\prod_{j=1}^{n}(\sigma_j^2)^{-\frac{1}{2}}exp(-\frac{1}{2}\sum_{j=1}^{n}\frac{(y_j-f(x_j))^2}{\sigma_j^2})\]

\subsection{Post-hoc draws of $u$}
For each MCMC iteration in aBCF, we draw a value of $u_j$ from its posterior conditional for each unit $j\in1,\dots,n$, as $u_j$ are independent of one another.

The prior for $u_j$ is simply: 
\[N(0,\sigma_u^2 )\]

The likelihood is:
\[N(y_j-f(x_j )|u_j,\frac{\sigma_\varepsilon^2}{w_j})\]

Yielding a conditional posterior of:
\[Pr(u_j|\dots) \propto exp(-\frac{u_j^2}{2\sigma_u^2}) exp(-\frac{w_j}{2\sigma_u^2}(y_j-f(x_j)-u_j)^2)\]

\subsection{Post-hoc draws of $u$ and $v$}
For each MCMC iteration in iBCF, we draw a pair of values of $u_j$ and $v_j$ from their joint posterior conditional for each unit $j\in 1,\dots n$, as each pair are independent across units.

The prior for the pair is multivariate normal:
\[
\begin{bmatrix}
	u_j \\ v_j
\end{bmatrix} 
\sim N(
\begin{bmatrix}
	0 \\ 0
\end{bmatrix} 
,
\begin{bmatrix}
	\sigma_u^2 & \rho\sigma_u\sigma_v \\
	\rho\sigma_u\sigma_v & \sigma_v^2 
\end{bmatrix} 
)\]

The likelihood is:
\[N(y_j-f(x_j) | 
\begin{bmatrix}
	1 & z_j
\end{bmatrix}
\begin{bmatrix}
	u_j \\ v_j
\end{bmatrix}
, \frac{\sigma_\varepsilon^2}{w_j})\]

Yielding a posterior conditional of:
\[Pr(
\begin{bmatrix}
	u_j \\ v_j
\end{bmatrix}
|\dots)
\propto
exp(-\frac{1}{2}
\begin{bmatrix}
	u_j & v_j
\end{bmatrix}
\begin{bmatrix}
	\sigma_u^2 & \rho\sigma_u\sigma_v \\
	\rho\sigma_u\sigma_v & \sigma_v^2 
\end{bmatrix} ^{-1}
\begin{bmatrix}
	u_j \\ v_j
\end{bmatrix}
)
exp(-\frac{w_j}{2\sigma_\varepsilon^2}(y_j-f(x_j)-u_j-z_jv_j)^2)
\]

\end{document}